\documentclass[preprint,showpacs,preprintnumbers,amsmath,amssymb,floatfix]{revtex4}    
    
\usepackage{graphicx}    
\usepackage{dcolumn}    
\usepackage{bm}    
\usepackage{rotating}    
\usepackage{epsfig}    
    
\begin{document}    
    
\title{Spin states of iron impurities in magnesium oxide under pressure: a possible  
intermediate state}    
    
\author{R. Larico$^{(1)}$, L. V. C. Assali$^{(2)}$, and J. F. Justo$^{(1)}$}    
    
\affiliation{    
$^{\rm (1)}$ Escola Polit\'ecnica, Universidade de S\~ao Paulo, \\    
CP 61548, CEP 05424-970, S\~ao Paulo, SP, Brazil \\    
$^{\rm (2)}$ Instituto de F\'{\i}sica, Universidade de S\~ao Paulo,\\    
CP 66318, CEP 05315-970, S\~ao Paulo, SP, Brazil}    
    
\date{\today}    
    
\begin{abstract}    
Ferropericlase, (Mg,Fe)O, is a major lower mantle mineral, and studying   
its properties is a fundamental step toward understanding the Earth's   
interior. Here, we performed a first principles investigation on the   
properties of iron as an isolated impurity in magnesium oxide, which is   
the condition of ferropericlase that iron-iron interactions   
could be neglected. The calculations were carried using the all-electron   
full-potential linearized augmented plane wave method, within the density   
functional theory/generalized gradient approximation plus the on-site Hubbard   
correction. We present the electronic and magnetic properties,  
electric and magnetic hyperfine splitting of this impurity in high and low spin   
states, for several charge states at zero pressure, which were then extended to   
high pressures. For the impurity in the neutral charge state, our   
results indicated that there is a metastable intermediate spin state (S=1), in  
addition to the high (S=2) and low (S=0) spin states.   
Those results were discussed in the context of an intermediate spin state,   
experimentally identified in ferrosilicate perovskite.    
\end{abstract}    
  
   
\maketitle    
  
\section{Introduction}    
\label{sec1}    
    
The identification of a pressure induced iron high-to-low spin transition in   
ferropericlase, $\rm (Mg_{1-x}Fe_x)O$ \cite{Badro}, and ferrosilicate   
perovskite, $\rm (Mg_{1-x}Fe_xSiO_3)$ \cite{Badro2}, has stimulated   
investigations on several aspects of this     
transition \cite{Lin05,Spez05,Gonch06,Tsuch06,Lin07,McCammon08,Wentz09,Wu09}.     
Since those are the two major minerals in Earth's lower mantle and the   
respective iron spin transitions were observed in mantle thermodynamic   
conditions, important geophysical implications could be anticipated, in   
terms of mantle chemical composition, heterogeneity, elasticity,   
and radiative transmission \cite{tac}.

In ferropericlase, an iron atom stays in a substitutional magnesium site,     
donating two valence electrons to its nearest neighboring oxygen atoms.     
The six remaining 3d-iron-related electrons generate a $t_2+e$ pair of   
orbitals in the crystalline field, with the $t_2$ orbital below the $e$ one.   
At low pressures, the exchange splitting prevails over the crystalline field   
one, favoring the high spin (HS) S = 2 state. With increasing pressure, the   
crystalline field splitting becomes more important, favoring the low spin   
(LS) S = 0 state beyond a certain transition pressure. Early experiments at   
room temperature \cite{Lin05} have suggested that this transition should be   
very sharp, occurring in a narrow pressure range. On the other hand, more   
recent theoretical \cite{Tsuch06,Wentz09,Wu09} and   
experimental \cite{Lin07,Crowh,marq2} investigations showed that it should   
be smoother, across wide pressure and temperature ranges. The current model   
for this transition describes ferropericlase as a solid solution with   
simultaneous concentrations of iron atoms in HS and LS states, which are   
determined by the thermodynamic conditions of the material \cite{Wentz09,Wu09}.

The phenomenology of iron in mantle minerals is very rich and a proper   
investigation on its properties is a fundamental step to build   
compositional models for the Earth's interior. However, there are many open   
questions that still need to be addressed, in order to understand the   
implications of the spin transition for the mantle physical properties.   
For example, there are several conflicting results in the literature   
concerning the elasticity of ferropericlase across this   
transition \cite{Crowh,marq2,marq,anto}. Another important element concerns   
the radiative conductivity of the mineral across the spin   
transition \cite{Gonch06,Keppler2007}, since such knowledge may help to   
build better temperature profiles of the Earth's lower mantle and core.   
There is also considerable uncertainty on the pressure range of that spin   
transition. While earlier experiments \cite{Badro,Lin05,Spez05,Gonch06}     
indicated that it occurred in the 30-40 GPa pressure range, at   
room temperature, more recent investigations suggested higher pressures   
for such transition \cite{Lin2010,Keppler2007,komabayashi2010}.

An equivalent high-to-low spin transition has been observed for iron in   
ferrosilicate perovskite \cite{Badro2,McCammon08}, the most abundant   
lower mantle mineral. An intriguing result is that while for iron in   
ferropericlase, only two spin states have been identified so far, in   
ferrosilicate perovskite, an intermediate (S=1) spin state has also been     
observed \cite{McCammon08}. This leads directly to the question whether this   
intermediate spin state could be also energetically favorable in   
ferropericlase. Several other questions are also open, such as the   
concentration of available electrons in the lower mantle, as result of
intrinsic or extrinsic defects in those minerals.   
This is an important property that could help   
to determine the charge state of iron atoms in minerals at those depths.   
Here, we present a theoretical investigation that explores some of those   
questions, considering the case of an isolated iron atom in magnesium oxide.   
In that case, iron was modeled as an impurity, such that the effects of   
iron-iron interactions could be neglected, and we could focus only on the   
properties of an isolated iron center. The first principles calculations were   
performed using the all-electron full-potential linearized augmented plane   
wave method, within the density functional theory and the generalized gradient   
approximation plus the on-site Hubbard correction in the 3d-related
iron states. The introduction of this   
potential, computed self-consistently, was a fundamental step to provide   
an appropriate description on the iron-related electronic energy levels.   
We computed the structural, electronic, and magnetic properties of iron in 
several charge states, and the respective electric and magnetic hyperfine  
splitting parameters of active centers. We then investigated the  
pressure effects on those properties, and compared our  
results to available experimental data.

\section{Methodology}    
\label{sec2}    
  
The calculations were performed within the density functional theory,  
using the all-electron full-potential linearized augmented plane wave (FP-LAPW)   
method, implemented in the WIEN2k package \cite{blaha}. The electronic exchange   
interaction was described within the generalized gradient approximation \cite{pbe}   
plus the Hubbard U potential correction (GGA+U) \cite{anisimov}. The on-site   
U values for the iron 3d-states were obtained   
self-consistently using the methodology described by Madsen and Nov\'ak \cite{novak}.

We considered a 54-atom MgO rocksalt reference supercell, in which an iron atom   
was placed in a substitutional magnesium site. The irreducible   
Brillouin zone was sampled by a grid of 2 $\times$ 2 $\times$ 2 k-points.     
Convergence on the total energy of the system was achieved using a 7.0/R parameter,     
which defines the total number of plane waves to describe the electronic wave   
functions in the interstitial regions, where R is the sphere radius of all the   
atomic regions (R = 0.90 \AA). For a certain atomic configuration, self-consistent   
iterations were performed until reaching convergence on both the total energy   
(10$^{-4}$ eV/atom) and the total charge in the atomic spheres   
(10$^{-5}$ electronic charges/atom). The positions of all atoms were relaxed, with   
no symmetry constrains, until the forces were smaller than 0.02 $\rm eV/\AA$ in   
any atom. In order to get results for different spin state configurations,     
some simulations were performed with constrained spin states.

The formation energy of a charged iron impurity in MgO,   
$\rm {E}_{f}^{q}(MgO:Fe_{Mg})$, is defined as \cite{assali2004}:    
\begin{eqnarray}    
\nonumber    
\rm {E}_{f}^{q}(MgO:Fe_{Mg})= & \rm { E_{tot}^{q} (MgO:Fe_{Mg})- E_{tot} (MgO)}    
\rm { + \mu_{Mg}- \mu_{Fe} + q (\epsilon_v + \varepsilon _F)}\ ,    
\label{eq1}    
\end{eqnarray}    
\noindent    
where $\rm E_{tot}^{q} (MgO:Fe_{Mg})$ is the total energy of a supercell,  
in the $\rm q$ charge state, containing the substitutional iron impurity,   
$\rm E_{tot} (MgO)$ is the total energy of a MgO crystal considering the same   
reference supercell. Additionally, $\epsilon_{\rm v}$ is the valence band top,   
adjusted to the band structures of the bulk material with and without the   
impurities, for each $\rm q$ charge state \cite{assali2004,assali2006}.   
This correction in the valence band top is necessary, due to  
inhomogeneities in the charge density of the finite primitive cell, which causes   
a Coulomb multipole interaction with its images, as discussed in   
ref. \cite{mattila}. Additionally, a uniform jellium background was implicitly   
considered to cancel out the long range multipole interactions of charged supercells  
\cite{assali2004}. $\rm {\varepsilon_F}$ is the Fermi level, $\mu _{\rm  Fe}$   
and $\mu _{\rm Mg}$ are respectively the Fe and Mg chemical potentials, computed   
within the same methodology described earlier, for the stable metallic   
crystalline phases.    
All the approximations and convergence criteria presented in the previous   
paragraphs have been shown to provide an accurate description on the electronic   
and structural properties of defect centers in a number of  
materials \cite{ayres2006,larico2009,assali2011}.

The introduction of a Hubbard potential correction represented a crucial   
element for an appropriate description of the 3d iron-related states in   
ferropericlase and iron oxide \cite{Tsuch06}, and consequently in the   
systems studied here. It is well established in the literature that the   
density functional theory provides a poor description of the strongly   
correlated electronic systems, such as the 3d-iron-related levels in the   
systems discussed here. The calculations with the local density  
or generalized gradient approximations lead to a metallic state for   
ferropericlase or iron oxide, although experimental results indicate that   
those systems are insulators \cite{Tsuch06}. The introduction of a Hubbard   
on-site correction increased the correlation interactions of 3d-related   
electronic levels, providing results more consistent with the available   
experimental data. There are several procedures to compute the Hubbard   
potential and, over the last few years, several U values have been used    
for those systems \cite{Tsuch06,persson}. On the other hand, our investigation  
computed this potential self-consistently \cite{novak}.

\begin{figure}[h]    
\begin{center}    
\includegraphics[width=100mm]{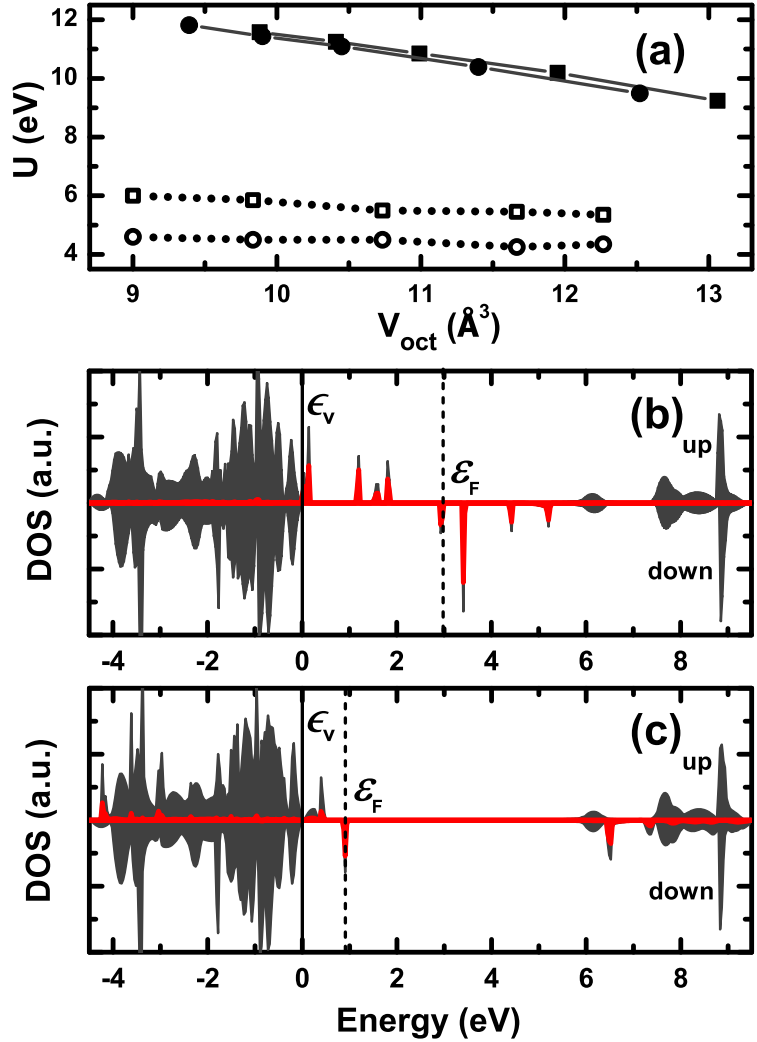}    
\caption{(a) Computed on-site Hubbard U values for the 3d states,   
as a function   
of the volume of the oxygen octahedral around the iron atoms, for HS (squares)   
and LS (circles) configurations. Our results (full symbols) are compared to those of   
other theoretical investigations (open symbols) \cite{Tsuch06}.    
(b) and (c) show the total density of states (TDOS) (dark gray regions) and partial   
density of states (PDOS) (solid lines in red) projected in the Fe 3d-related  energy   
levels of the HS state, without and with the correction, respectively.   
The spin up (down) states are represented in top (bottom) of the figs. (b)   
and (c). Additionally, the energy reference is set at the valence band top   
of MgO ($\epsilon_v = 0$), and the dashed lines represents the highest 
occupied level.}    
\label{fig1}    
\end{center}    
\end{figure}

Figure \ref{fig1}(a) shows the values of the Hubbard potential as a function   
of the volume of the oxygen octahedron around the iron atom, as compared   
to the values used by other authors for ferropericlase. Only for reference,   
some authors used constant values (3 or 5 eV) for the Hubbard potential   
irrespective of the octahedral volume \cite{persson}. The self-consistent values   
used here are considerably larger than the ones from other calculations. However,
it should be pointed out that the value of this potential depends   
strongly on the methodology: while our calculations  
were performed within the full potential method (FP-LAPW), others were   
performed within the pseudopotential method \cite{coco}. For example,
in the $\rm Fe_2O_3$ system, the iron Hubbard potential  
was found to be 8.73 eV within a full potential calculation \cite{novak} and  
3.3 eV within a pseudopotential calculation \cite{blan}.
Additionally, other recent theoretical investigations  
have also used large U values for the 3d-related energy levels of   
iron \cite{novak,nabi,ibrahim}, consistent with our values.

Figure \ref{fig1}(b) and (c) show respectively the density   
of states of the system without  and with the correction, indicating   
that it provides an appropriate description of the electronic   
structure. Without the correction, the energy difference between the highest   
occupied and lowest unoccupied  iron-related levels was 0.5 eV. This energy   
difference changed to 5.5 eV with the correction.

\section{Results}    
\label{sec3}    
    
\subsection{Zero Pressure Results}  
  
Under ambient conditions, MgO crystallizes in the rocksalt (B1) structure,   
with a measured lattice parameter of $a_{expt}$ = 4.216 ${\rm \AA}$, presenting   
a large direct electronic bandgap of 7.67 eV \cite{landolt-livro}. In this   
study, using the approximations presented in the previous section, we found an   
equilibrium lattice parameter of $a_{th}$ = 4.21 ${\rm \AA}$ and a direct   
bandgap of E$_g$ = 4.50 eV. Those values are consistent with values from other   
theoretical investigations using similar approximations \cite{karki} and in   
good agreement with experimental data \cite{landolt-livro}.

We considered several charge and spin states for the substitutional iron   
impurity, using the supercell with the MgO theoretical lattice parameter.   
We computed the formation and transition energies, along with the respective   
structural and electronic properties of HS and LS states. For a certain   
charge state, in order to build an energy stability curve as a function of   
iron magnetic moments, we performed a set of calculations constraining the  
total spin of the system.

Table \ref{tab1} presents the properties of the impurity in  several $\rm q$   
charge states, $\rm (MgO:Fe_{Mg})^{q}$. The presence of a substitutional Fe   
impurity in MgO caused important relaxations on the neighboring oxygen atoms.   
For the impurity in the neutral charge state in its HS, the system presented   
an outward relaxation with respect to the original MgO crystalline  structure.   
The volume of the oxygen octahedron around a Mg atom was 12.81 \AA$^3$ in MgO,   
while with the substitutional iron impurity in its HS state it changed to   
13.45 \AA$^3$. This outward relaxation hides an important symmetry lowering,   
due to a Jahn-Teller distortion, as result of the iron-related partially   
occupied electronic energy levels. Going from this HS state to the LS one, there   
was a substantial inward relaxation, with the volume of the oxygen octahedron   
going to 13.08 \AA$^3$, representing a 3\% volume reduction with respect to that   
in the HS state. Additionally, since the LS state presented a full shell   
electronic occupation, it had an octahedral symmetry. This volume reduction was   
in reasonably good agreement with the 7\% reduction, computed in another   
theoretical investigation on ferropericlase ($\rm Mg_{1-x}Fe_x O$ with   
$\rm x = 0.1875$ \cite{Tsuch06}. The calculations for that alloy allowed   
crystalline relaxation with the presence of Fe, while our   
calculations considered a fixed MgO lattice, justifying a smaller inward relaxation   
on the octahedron. According to table \ref{tab1}, for the positive and doubly   
positive charge states, there were also substantial inward relaxations going   
from a HS to a LS state. Additionally, for a certain spin state (HS or LS),   
the octahedron became smaller as going from a neutral charge state to  
positive or doubly positive ones.

We found that Fe was stable only in three charge states (neutral, positive, and   
doubly positive), presenting 3d-related energy levels in the MgO bandgap.   
For the neutral charge state, we computed the properties of the HS and LS states,   
and found that the formation energy of the HS state is 
1.44 eV lower than that of the LS one. Consistent with    
theoretical \cite{Tsuch06,Wentz09} and experimental \cite{Lin05,Spez05}     
results for ferropericlase, our results indicated that the HS state is considerably   
more stable than the LS one at low pressures. For the positive charge state, there   
are also the HS and LS states, corresponding to 5/2 and 1/2 spin values.    
Here, the formation energy of the HS state is 2.12 eV lower than the LS   
one. For the doubly positive charge state, there are also the HS and LS states,   
respectively with S=2 
and S=1, with the HS state having lower formation energy than the LS one. The   
results indicated that, for low pressures, the HS state was more stable than the    
LS one, irrespective of the charge state of the center.

\begin{figure}[h]    
\begin{center}    
\includegraphics[width=120mm]{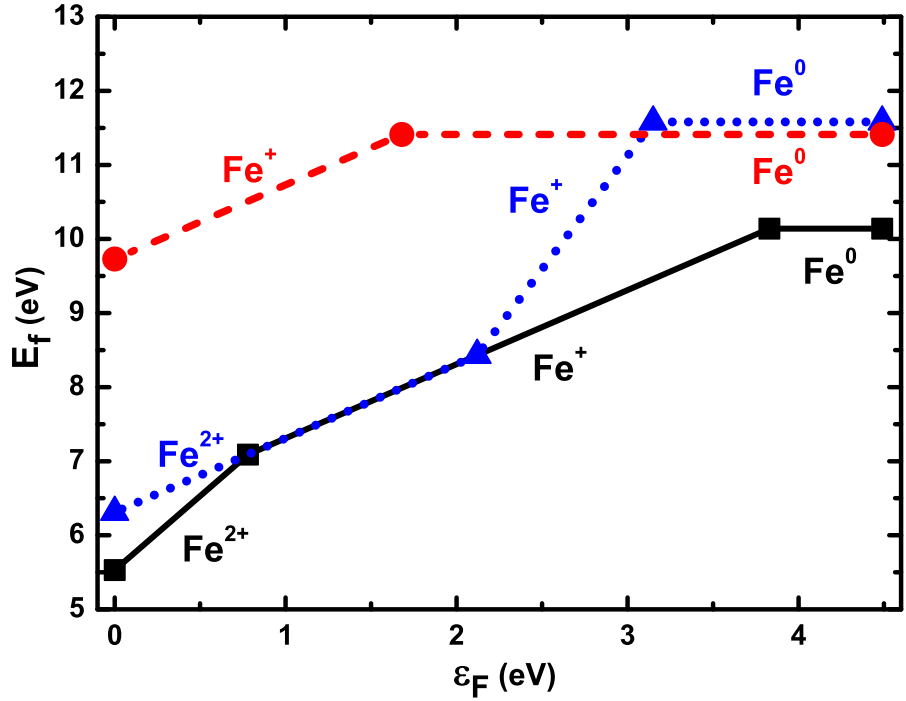}    
\caption{Formation energies of (MgO:Fe)$^{\rm q}$ as a function of Fermi energy   
position, in the MgO bandgap, for the isolated impurity. The zero of   
Fermi-level corresponds to the valence band top. The   
Square, triangle, and circle symbols  
correspond respectively to HS, LS, and IS states at zero pressure.}    
\label{fig2}    
\end{center}    
\end{figure}

Figure \ref{fig2} shows the formation energy of the $\rm (MgO:Fe_{Mg})^{q}$     
center at several $q$ charge states as a function of the MgO Fermi 
level ($\rm 0 \le {\varepsilon _F}  \le E_g$), where the valence band top
was set to zero ($\rm {\epsilon _v} = 0$) and $\rm E_g$ is the materials band gap.     
For the HS state, the doubly positive charge state is stable for   
$\rm 0 \le {\varepsilon _F} \le $ 0.78 eV, the positive charge state is stable for    
0.78 $\le {\varepsilon _F} \le $ 3.83 eV, while the neutral   
one is stable for $\rm {\varepsilon _F} >$ 3.83 eV.     
According to figure \ref{fig2}, since formation energy of the LS state is   
higher than the one of the HS state, the stability curve corresponding to the   
LS state is over the curve corresponding to the HS one for any value of the   
Fermi level. Since the density functional theory calculations underestimate   
the MgO bandgap by more than 3 eV, when compared to experimental   
data \cite{landolt-livro}, as discussed earlier, then the neutral charge   
state, $\rm (MgO:Fe_{Mg})^{0}$, is expected to be the stable configuration   
for the Fermi level lying in the top half part of the MgO bandgap.

\begin{figure}[h]    
\begin{center}    
\includegraphics[width=120mm]{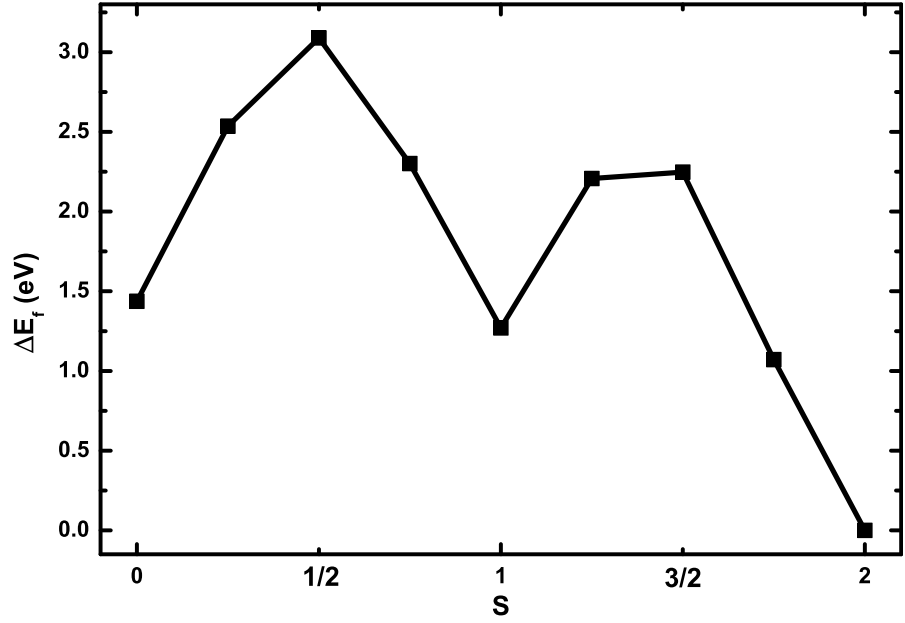}    
\caption{Total energy difference ($\rm \Delta E$), with  
respect to the total energy of the HS ground state, of   
$\rm (MgO:Fe_{Mg})^0 $ as a function of the spin   
of the center (S) at zero pressure.   
The solid lines are only a guide to the eyes.}    
\label{fig3}    
\end{center}    
\end{figure}

We also computed the energy barrier, as a function of the total magnetic moment,   
for the systems going from a HS toward a LS state.  In order to perform such   
analysis, the energetics of the systems were computed with constrained   
intermediate magnetic moments between the HS and LS configurations, in the   
neutral charge state. Fig. \ref{fig3} presents the total energy, of the neutral   
charge state, as a function of spin configurations of the centers between S=0   
and S=2 states. The results indicated that going from a HS state to a LS one,   
there was a metastable intermediate spin (IS) state, with S=1.   
The Hubbard potential value of the IS state is essentially the same of the  
HS one. This IS state had  
a total energy of about 1.30 eV higher than that of the HS state, and even a   
little smaller than the LS one. This IS state had  
a volume of the oxygen octahedron of 13.30 ${\rm \AA}^3$, which is a   
little smaller than that of the HS state (of 13.45 ${\rm \AA}^3$).   
The presence of an intermediate spin state was carefully explored in order
to check if it was not only a theoretical artifact. In order to explore this
IS state, we performed calculations with several U values (from 0 to 9 eV),
and we found that the IS state is metastable in all those cases. Additionally,
we performed calculations with unconstrained magnetic moments around
S=1 (with S=0.75 and S=1.25), and both calculations converged toward the 
IS state with S=1.  For the positive charge state, we  
also observed an IS state in a metastable configuration, while  
for the doubly positive one, we found no IS state.

At zero pressure, the large difference   
in total energies, for the neutral charge state, indicated that essentially all   
the iron centers would be in the HS state. Considering the solid solution model   
with concentrations of HS and  LS states \cite{Wentz09,Wu09}, our results  
indicated that there would be very small concentrations of LS and IS states at   
low pressures \cite{kantor}, even at high temperatures. The energy barriers for the spin   
crossover are very large for both the HS-IS and IS-LS transitions.   
Here, we should point out that the introduction of a Hubbard potential,
to provide a proper description of the electronic structure of the system,
comes with a price: a poorer description of the total energy of the system.
Several attempts have been introduced in the literature to improve the
description of the total energy \cite{jain}. Therefore, there are important
uncertainties in the energy barriers for the spin crossover.

We also found that the IS configuration is a possible spin state for Fe in 
ferropericlase, and is   
fully consistent with the experimental observation of an equivalent IS state   
in ferrosilicate perovskite \cite{McCammon08}. However, the small difference   
in the oxygen octahedron volume, between the HS and IS states, may hamper an   
identification of this center in ferropericlase using several experimental  
methodologies. A possible way to identify such center is by electron   
paramagnetic resonance (EPR) spectroscopy, in which the hyperfine parameters   
could be measured. Table \ref{tab1} presents the magnetic hyperfine parameters for all   
the electrically active iron centers in MgO at zero pressure.   
For  the $\rm (MgO:Fe_{Mg})^{0} $ center,  
the differences in those parameters for HS and IS states may allow an identification   
of this IS state by EPR measurements.

\begin{figure}[h]    
\begin{center}    
\includegraphics[width=140mm]{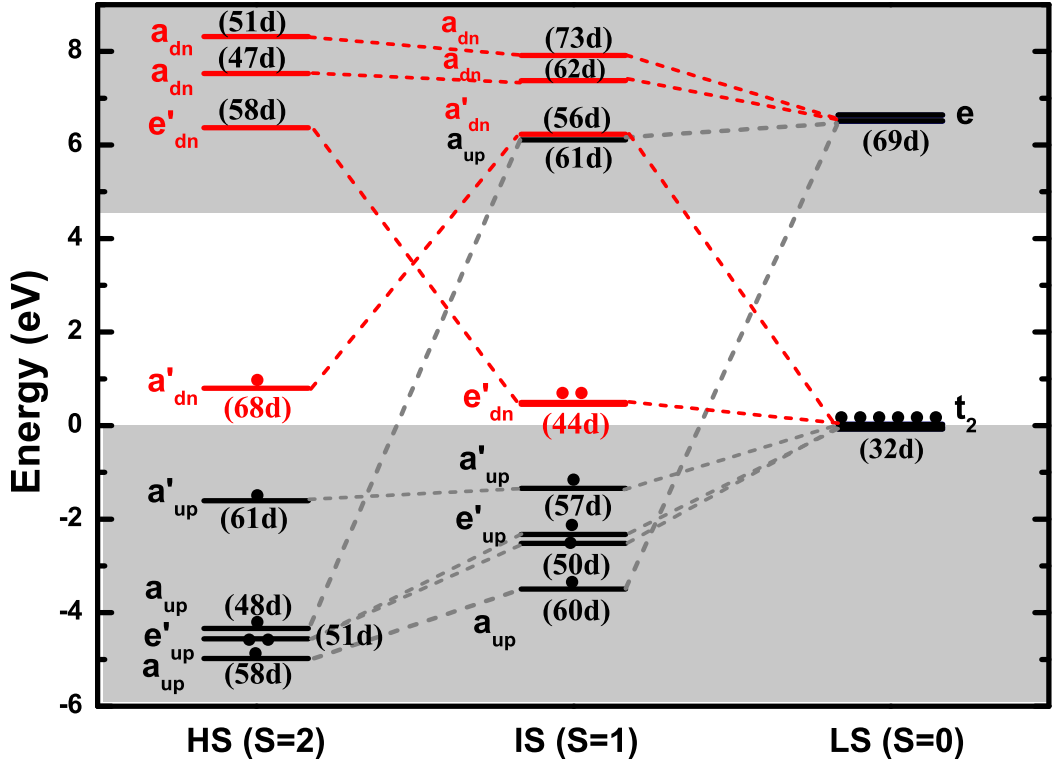}    
\caption{The Kohn-Sham spin-polarized energy eigenvalues around the   
$\Gamma$ point, representing the 3d-related Fe levels around the gap region   
for isolated substitutional neutral Fe in several spin configurations:   
(a) S=2 (HS), (b) S=1 (IS), and (c) S=0 (LS). The gap level occupations are   
given by the numbers of filled circles. Numbers in parentheses represent the   
d-character percentage of charge inside the Fe atomic sphere. Up and down   
arrows represent the spin up and spin down levels, respectively.}    
\label{fig4}    
\end{center}    
\end{figure}

The difference in the electronic structure between the HS and IS states  
could also help identifying such IS state.  
Figure \ref{fig4} presents the electronic structure of the 3d-iron-related energy   
levels in $\rm (MgO:Fe_{Mg})^{0}$ for     
the HS, IS, and LS states. In the LS state, the system consists of only   
$t_2+e$ levels in a close shell configuration. The $t_2$ level is fully   
occupied near the valence band top of MgO and the $e$ one is unoccupied in the   
conduction band. For the HS state, due to the symmetry lowering, in comparison   
to the LS state, the $t_2$ level splits into $a+e$ levels. The HS state consists   
of five electrons with spin up and one with spin down. The last occupied level   
is an $a_{\downarrow}$ orbital, with 69\% of d-character (inside the
Fe sphere), about 1 eV higher   
than the valence band top of MgO. Finally, the IS state has the highest occupied   
level as an $e_{\downarrow}$ with two electrons,  and with 48\% of d-character.   
This level is near the valence band top, while the first unoccupied level is   
inside the conduction band.

\subsection{High Pressure Results}  
  
A possible way to explore the properties of iron impurities in MgO  
is observing the pressure effects. This would be specially important  
to identify the IS state, using EPR or M\"osbauer spectroscopies.   
  
First of all, in order to provide consistent results for MgO under pressure, we   
explored the elastic properties of MgO. Our calculations indicated a bulk   
modulus of 152 GPa, using a third-order Birch-Murnaghan equation of state,   
which is fully consistent with the values obtained by other theoretical   
investigations using similar approximations \cite{karki} and in excellent   
agreement with experimental data \cite{landolt-livro}. This equation of   
state was later used to obtain the dependence of the MgO lattice parameter   
with external pressure.

\begin{figure}[h]    
\begin{center}    
\includegraphics[width=140mm]{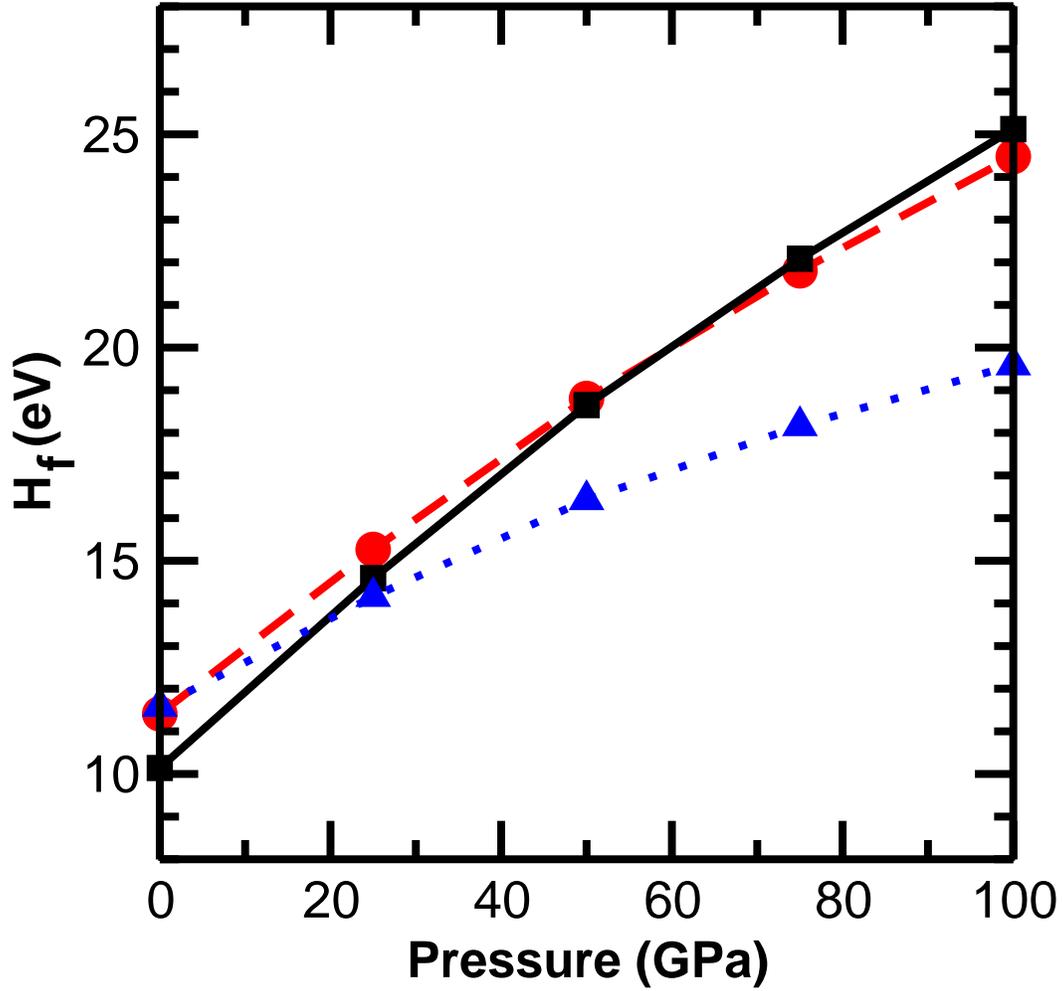}    
\caption{Enthalpy of formation  ($\rm H_f$) of iron impurity in magnesium oxide 
in the neutral charge state,  
$\rm (MgO:Fe_{Mg})^0 $, as a function of pressure at different spin states. 
Square, triangle, and circle symbols  
correspond respectively to HS, LS, and IS states.}  
\label{fig5}    
\end{center}    
\end{figure}

An important property is to check the stability of 
the spin states as a function of pressure. In ferropericlase, high to low
spin transition has been observed in the 30-50 GPa range at room temperature, depending on the 
concentration of iron in the material \cite{fei,yoshino}. For very low
iron concentrations, it has been found a transition near 30 GPa.
In order to check the stability of the spin states of iron atoms in MgO, 
we computed the respective enthalpy
of formation as a function of pressure for different spin states at the neutral charge state. 
In order to obtain the enthalpy of formation, we computed the respective 
volumes of formation, following the procedure used in other systems \cite{centoni}. 
Figure \ref{fig5} shows the enthalpy of formation as a function of pressure for
iron in MgO. The results indicated that the HS and LS curves cross each other at about 20 GPa,
indicating an spin transition. This result is fully consistent with experimental
data for ferropericlase at low iron concentrations \cite{yoshino}. The figure also shows that
across the pressure range, the IS is always higher in energy than other spin states.

We observed that, for all charge states, the external pressure does not   
modify the point symmetry of the centers, only compressing the respective oxygen   
octahedra. Therefore, pressure modifies only slightly the center structures,  
affecting more strongly the electronic structures, as the energy levels  
are shifted with respect of the materials band gap.

\begin{figure}[h]    
\begin{center}    
\includegraphics[width=140mm]{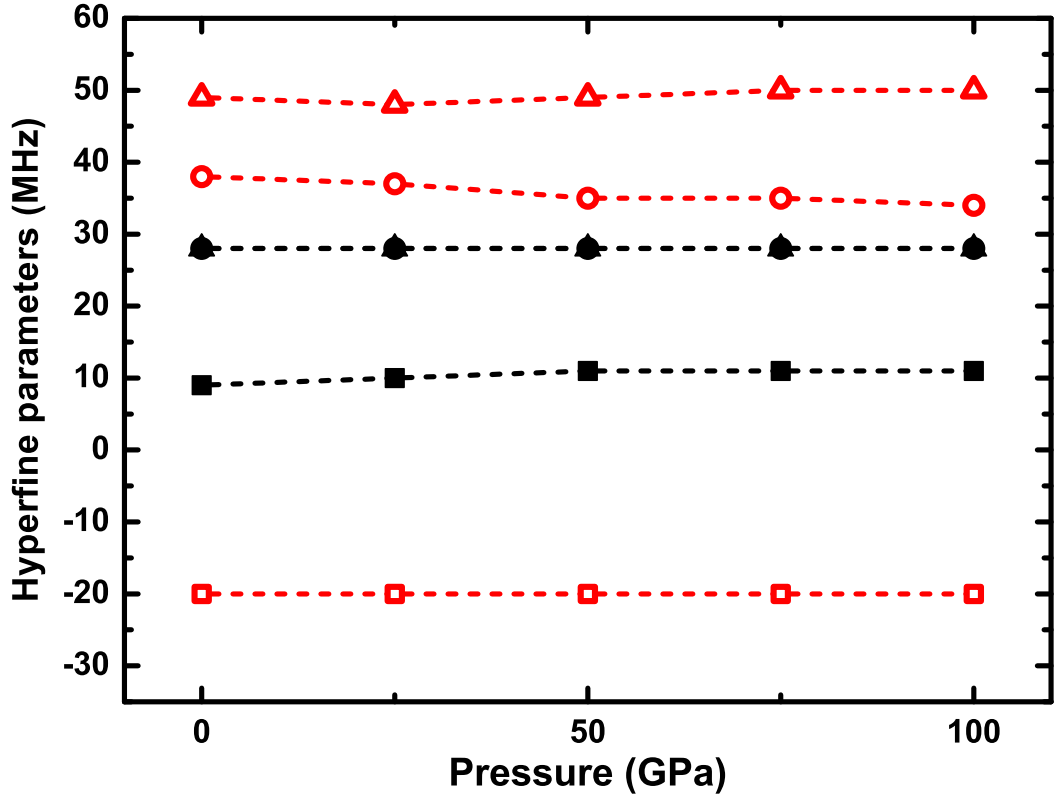}    
\caption{Magnetic hyperfine parameters: A$_{1}$ (square), A$_{2}$ (triangle), and A$_{3}$   
(circle) for the $\rm (MgO:Fe_{Mg})^{0} $ center, in HS (close symbols) and IS   
(open symbols) states, as a function of pressure.  The pressure, as a function  
of the lattice parameter, was computed  
considering the fitting of a third-order Birch-Murnaghan equation of state.}  
\label{fig6}    
\end{center}    
\end{figure}

Figure \ref{fig6} presents the theoretical magnetic hyperfine parameters for the   
HS and IS states in the neutral charge state as a function of pressure.   
The large differences in the hyperfine parameters, between HS and IS states,   
may allow a proper identification of this IS state in ferropericlase. The   
figure also shows that those parameters are essentially insensitive to   
external pressure.

\begin{figure}[h]    
\begin{center}    
\includegraphics[width=130mm]{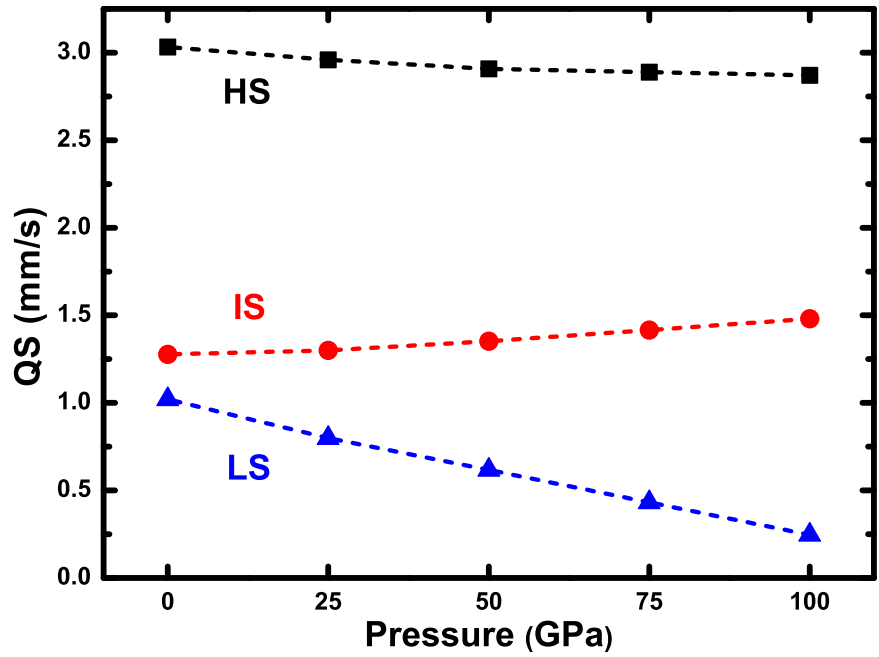}    
\caption{Quadrupole splitting (QS) of $\rm (MgO:Fe_{Mg})^{0}$ center   
in HS (squares), IS (circles) and LS (triangles) states   
as a function of pressure.}   
\label{fig7}    
\end{center}    
\end{figure}

Nuclear quadrupole resonance is another technique that could allow to    
distinguish the HS and IS states, in which the measurements are associated to   
the electric field gradient (EFG) of each center. We computed the EFG of all   
spin states in the  $\rm (MgO:Fe_{Mg})^{0}$ as a function of pressure. The   
computed electric field gradient, $\rm V_{zz}$, at the center of the iron   
nucleus is converted to the quadrupole splitting (QS) value, using  the relation   
$\rm QS \simeq eQV_{zz}/2h$, where $\rm (h/e)=4.1356692\times  10^{-15}[V/MHz]$   
and $\rm Q$ denotes the nuclear electric quadrupole moment of iron. The EFGs   
were converted to the QS values using  the $^{\rm 57}$Fe nuclear quadrupole moment   
of $\rm Q=0.16 \pm 0.02 \ barn$ ($\rm 1 barn = 10^{-28} m^2$) \cite{petrilli}.  
Figure \ref{fig7} shows the QS of the $\rm (MgO:Fe_{Mg})^{0}$ center, as a   
function of pressure, in the HS, IS, and LS configurations. The QS is very
sensitive to increasing external pressure, decreasing for the HS state, 
while increasing for the IS one, which  may help to distinguish the HS and IS   
states.

\begin{figure}[h]    
\begin{center}    
\includegraphics[width=130mm]{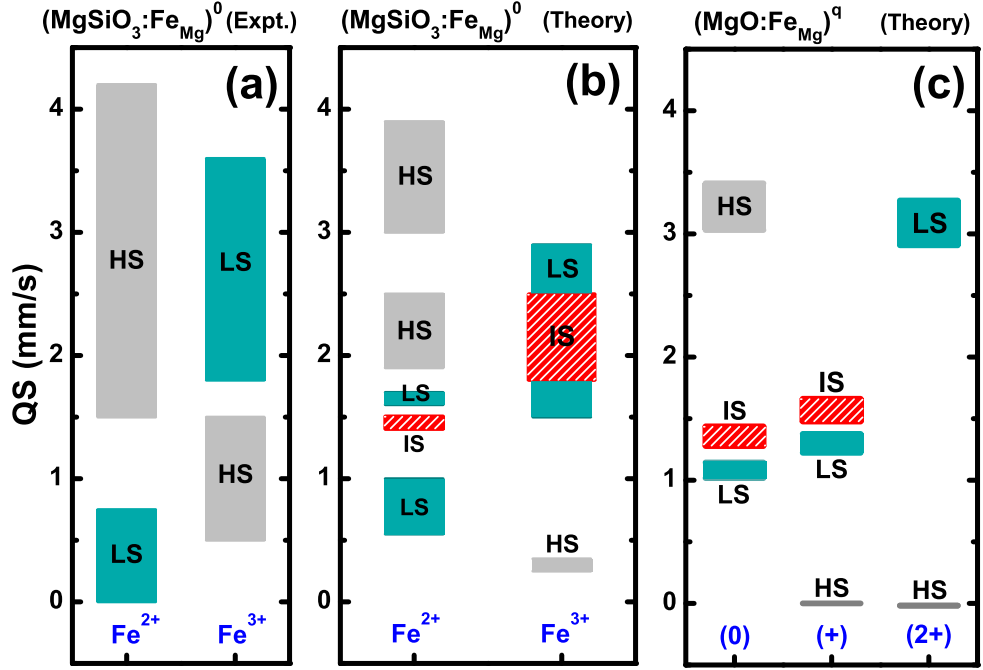}    
\caption{Our theoretical results (c) for the QS in   
$\rm (MgO:Fe_{Mg})^{q}$ (q = 0, +, or 2+)   
for zero pressure, compared to experimental   
\cite{lin2011} (a) and theoretical \cite{hsu2011} (b)   
QS values at the Fe atom in ferrosilicate perovskite, $\rm (MgSiO_3:Fe_{Mg})^{0}$.   
In ferrosilicate  perovskite, Fe can have 2+ or 3+ oxidation charge states.}   
\label{fig8}    
\end{center}    
\end{figure}

Figure \ref{fig8} compares the theoretical QS values for the 
$\rm (MgO:Fe_{Mg})^{q}$, in several charge states with the values of an iron  
atom in ferrosilicate perovskite, represented as the $\rm
(MgSiO_3:Fe_{Mg})^{0}$ center in the figure, obtained by recent   
theoretical \cite{hsu2011} and experimental \cite{lin2011} investigations.  
Ferrosilicate perovskite has two different sites for iron. As a result,  
for the $\rm (MgSiO_3:Fe_{Mg})$ center, in the neutral charge state,  
the iron atom can stay in two net charge states (2+ or 3+), depending on the 
lattice site, meaning that iron can donate two or three  electrons to its 
neighboring oxygen atoms.  According to the figure, the 
$\rm (MgO:Fe_{Mg})^{0}$ center has QS values, for all three spin states, that are 
consistent with the respective ones associated to the iron in a 2+ oxidation 
state ($\rm Fe^{2+}$)  in ferrosilicate perovskite. This provides an
additional confirmation  that the iron atom in MgO in a neutral charge state,   
$\rm (MgO:Fe_{Mg})^{0}$, has a 2+ oxidation charge state, i.e., it donates 
two electrons to its neighboring oxygen atoms. On the other hand, the   
$\rm (MgO:Fe_{Mg})^{+}$ center has QS values, for all three spin states, 
that are consistent with the  respective ones  associated to the iron 
$\rm Fe^{3+}$ in ferrosilicate perovskite, suggesting that same oxidation 
charge state for iron in MgO. Finally, the $\rm (MgO:Fe_{Mg})^{2+}$ center 
has QS values  for spin states that resemble the ones of the  
$\rm (MgO:Fe_{Mg})^{+}$  center. This suggests that iron in the 
$\rm (MgO:Fe_{Mg})^{2+}$ center has a 3+ oxidation charge state, and not 
a 4+, as it could be expected by a simple inspection. All those results 
indicate similarities in the properties  of iron atoms in  
MgO or ferropericlase with those in ferrosilicate perovskite.  
Such similarities of iron in those two materials are not fortuitous, they indicate 
that the d-related (iron) orbital occupancy plays a major role on the
properties of those centers, even more important than the center symmetry or 
the number of neighboring oxygen atoms.

\section{Summary}    
\label{sec4}    
    
In summary, we performed a first principles investigation on the properties    
of substitutional iron impurities in magnesium oxide. We found that those
centers can  stay stable in three charge states: neutral, positive and doubly 
positive ones. The center in neutral charge state controls the top part of the 
band gap in HS and LS states. For the lower mantle   
properties, in which  iron is incorporated in ferropericlase alloys, this   
property should be discussed, since the Fermi level, which depends on the amount    
of available carriers in the system, is determined by the concentrations    
of other intrinsic and extrinsic defects, such as vacancies, interstitials, 
dislocations, and impurities.     
    
We also found that the controlling mechanism of spin transition is associated   
to the energy barrier between the different spin states. Additionally, we found   
that there are three possible spin states for the iron in neutral charge state,   
consistent with results for the iron spin states in ferrosilicate perovskite.   
The intermediate spin state, with S=1, is a metastable configuration that could   
be observed in MgO in minor concentrations by EPR or M\"osbauer spectroscopy.   
Additionally, our calculations on the quadrupole splitting indicated similarities  
of iron atoms in ferropericlase and ferrosilicate perovskite.  
All those aspects   
should be discussed in the context of the studies on the charge and radiative   
conductivity of the lower mantle, with important geophysical implications for   
the temperature profile of the inner layers of the Earth.

\section*{Acknowledgments}    
The authors acknowledge partial support from Brazilian agencies FAPESP and CNPq.    
The authors thank Prof. W. V. M. Machado for fruitful discussions.

\pagebreak

\begin{table}[h]    
\caption{Results for isolated substitutional Fe impurities in MgO at zero pressure:   
spin state   
configuration, total spin (S), the localized magnetic moment inside the atomic   
iron sphere ($\rm \mu_B^{Fe}$), the volume of the oxygen octahedron around the   
Fe atom (V$_{\rm oct}$), formation (E$_{\rm f}$) and transition (E$_{\rm t}$) energies   
for several $q$ charge states, and magnetic hyperfine parameters (A$_{1}$, A$_{2}$, and A$_{3}$).    
The transition energies were computed with respect to the MgO valence band top   
($\rm \epsilon_v$), as discussed in the text. Volumes, magnetic   
moments, energies, and hyperfine parameters are given respectively in \AA$^3$,  
Bohr magneton, eV, and MHz.}    
\label{tab1}    
\vspace{0.0cm}    
\begin{center}    
\begin{tabular}{lccccccccccc}    
\hline \hline    
\vspace{0.00mm} \\    
{~Center~}  &{~~State~~}&{~~S~~}&{~~$\rm {\mu_{B}^{Fe}}$~~}& V$_{\rm oct}$ &  
{~~~~~~~~~E$_{\rm f}$~~~~~~~~~}&    
{~~~~~~~~~E$_{\rm t}$~~~~~~~~~} & A$_{1}$ & A$_{2}$ & A$_{3}$ \\    
\hline    

\vspace{0.001mm} \\    
$\rm (MgO:Fe_{\rm Mg})^{0}$  &  HS &  2       &     3.44       & 13.45 &   
10.14 &   & 9 & 28 & 28  \\    
                &  IS &   1       &     1.73       & 13.30 &   
11.41 &     & -20 & 49 & 38  \\    
                &  LS &  0       &     0.0         & 13.08 &  
11.58  &    &  0 & 0 & 0 \\    
\vspace{0.001mm} \\    
$\rm (MgO:Fe_{Mg})^{+}$  &  HS & ~~5/2~~  &  4.13  & 12.02 &   
6.31 + $\varepsilon_F$ & 3.83  (+/0) & ~20~ & ~20~ &  ~20~ \\    
                       & IS  &   3/2    &  2.80      &  11.67  &  
9.73 + $\varepsilon_F$ & 1.68   (+/0) & -7  &  34    & 34 \\  
                       &  LS &   1/2     &  0.98       & 11.20 &   
8.43 + $\varepsilon_F$ & 3.15  (+/0) & 4 & 36 & 22 \\    
\vspace{0.00mm} \\    
$\rm (MgO:Fe_{Mg})^{2+}$ &  HS &   2       &     ~~4.12~~       & ~~11.94~~ &  
5.53 + 2$\varepsilon_F$& 0.78 (2+/+) & 25  & 25  &  25 \\    
                &  LS &   1   &  2.55   & 10.45 &   
6.31 + 2$\varepsilon_F$& 2.12 (2+/+) &  -5 &  38 & 38      
\vspace{0.1mm} \\    
\hline \hline    
\end{tabular}    
\end{center}    
\end{table}    
\pagebreak

\newpage    
 \pagebreak

\pagebreak

\pagebreak

\pagebreak

\pagebreak

\pagebreak

\pagebreak

\pagebreak

\end{document}